\documentclass{ws-procs9x6}            % default, citations in superscript
\begin{document}
\title{Accretion disks around a mass with quadrupole}

\author{M. Abishev, K. Boshkayev, H. Quevedo and S. Toktarbay$^*$}

\address{Physical-Technical Faculty, Al-Farabi Kazakh National University,\\
Al Farabi av. 71, 050040 Almaty, Kazakhstan\\
 Instituto de Ciencias Nucleares, 
Universidad Nacional Aut\'onoma de M\'exico, \\
 AP 70543, M\'exico, DF 04510, Mexico\\ 
$^*$E-mail: saken.yan@yandex.com}

\begin{abstract}
We consider the stability properties of test particles moving along circular orbits
around a mass with quadrupole. We show that the quadrupole modifies drastically 
the properties of an accretion disk made of such test particles. 
\end{abstract}


\bodymatter

\ 

%\section* %{The $q-$metric and accretion disks}

The simplest generalization of the Schwarzschild metric which contains a quadrupole parameter $q$ is given by
\begin{eqnarray}
ds^2 & = & \left(1-\frac{2m}{r}\right) ^{1+q} dt^2  - \left(1-\frac{2m}{r}\right)  ^{-q}\nonumber\\
&&\left[ \left(1+\frac{m^2\sin^2\theta}{r^2-2mr }\right)^{-q(2+q)} \left(\frac{dr^2}{1-\frac{2m}{r} }+ r^2d\theta^2\right) + r^2 \sin^2\theta d\varphi^2\right]  .
\label{zv}
\end{eqnarray}
This solution is known as the  $\delta-$metric or as the $\gamma-$metric and  was first obtained by Zipoy and Voorhees \cite{zv68}. 
We propose to use the term  quadrupole metric ($q-$metric) to emphasize the role of the parameter $q$. 
The $q-$metric is an axially symmetric exact vacuum solution, and reduces to the Schwarzschild metric for $q\rightarrow 0$. 
It is asymptotically flat with a central curvature singularity at $r=0$ and an outer singularity at $r=2m$ which is naked.
According to the Geroch definition, the independent multipole moments are the monopole $M_0=m(1+q)$ and the quadrupole 
$M_2=-\frac{m^3}{3}q(1+q)(2+q)$. For more details, see Ref. \citenum{quev11}.

\begin{figure}%
\includegraphics[scale=0.11]{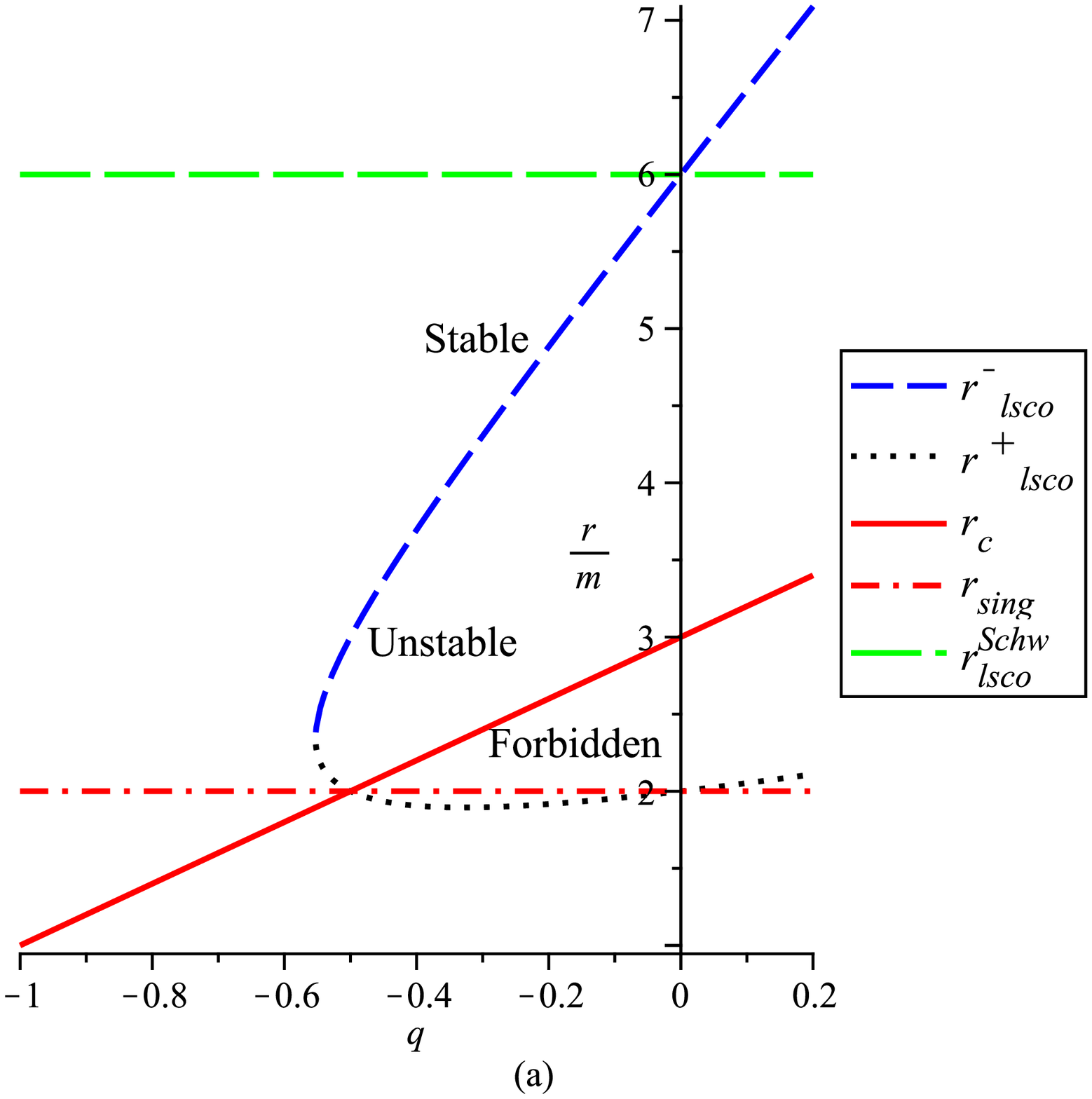} \qquad
\includegraphics[scale=0.18]{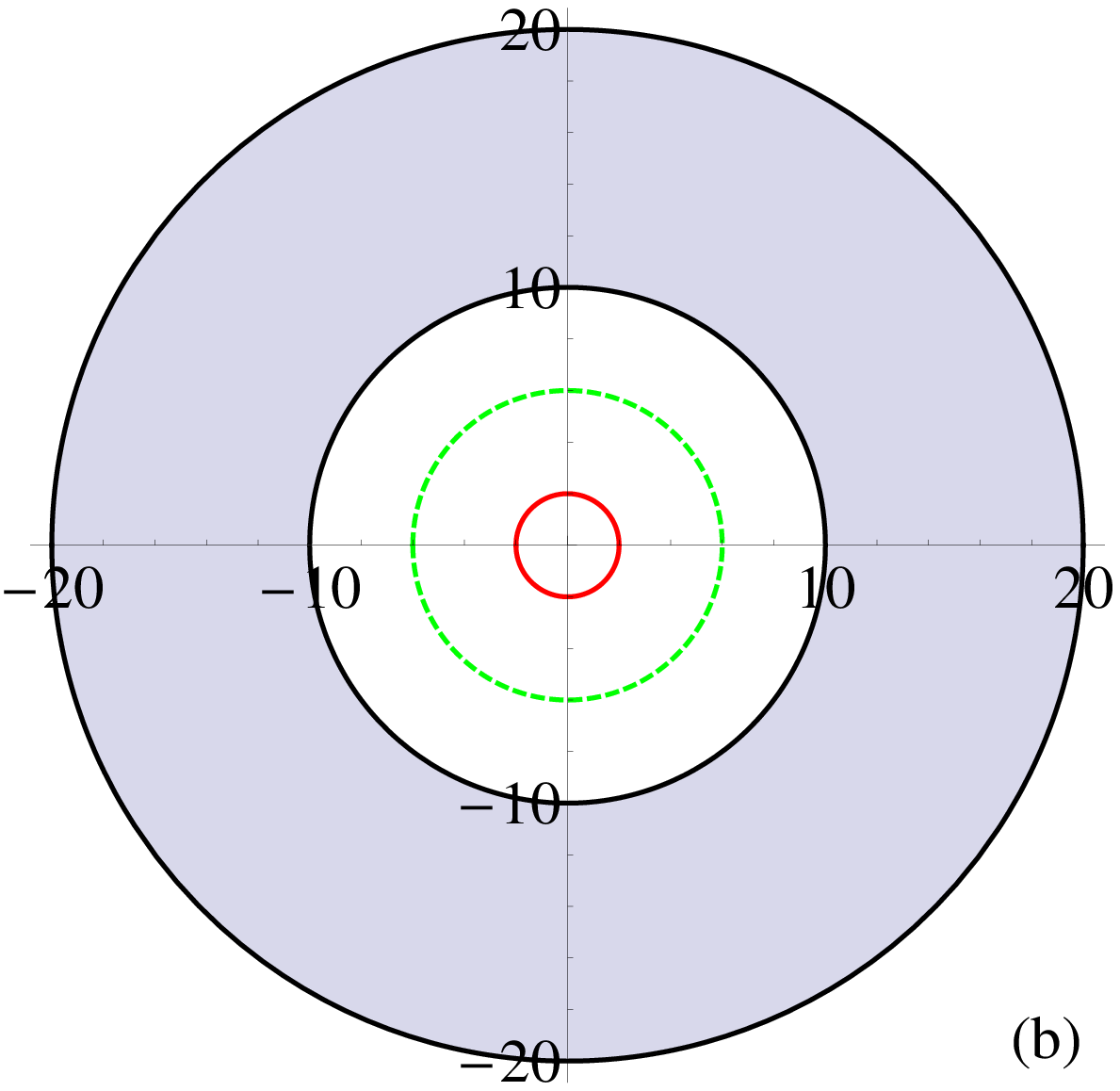}  \qquad
\includegraphics[scale=0.18]{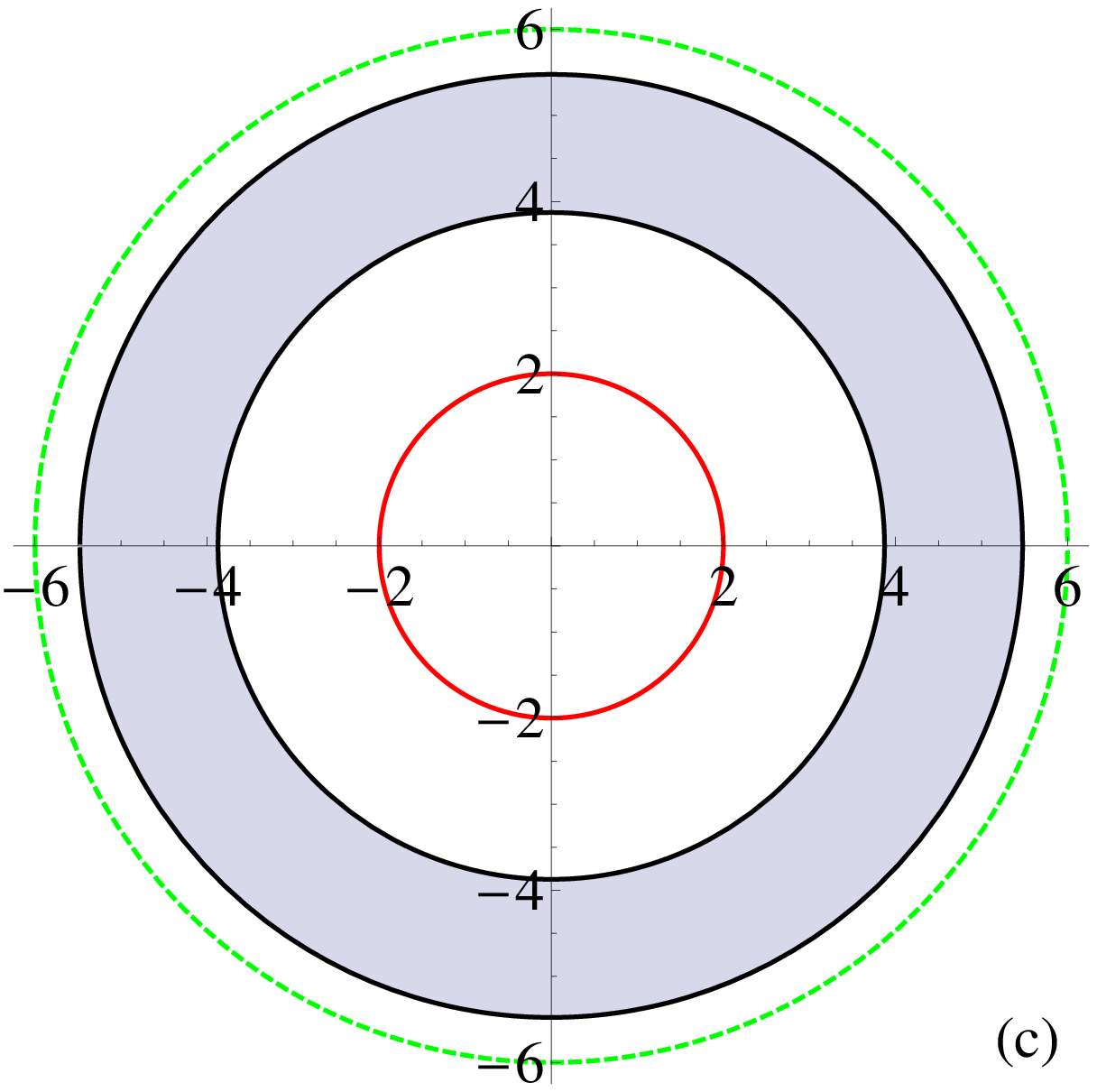}  \qquad
\includegraphics[scale=0.18]{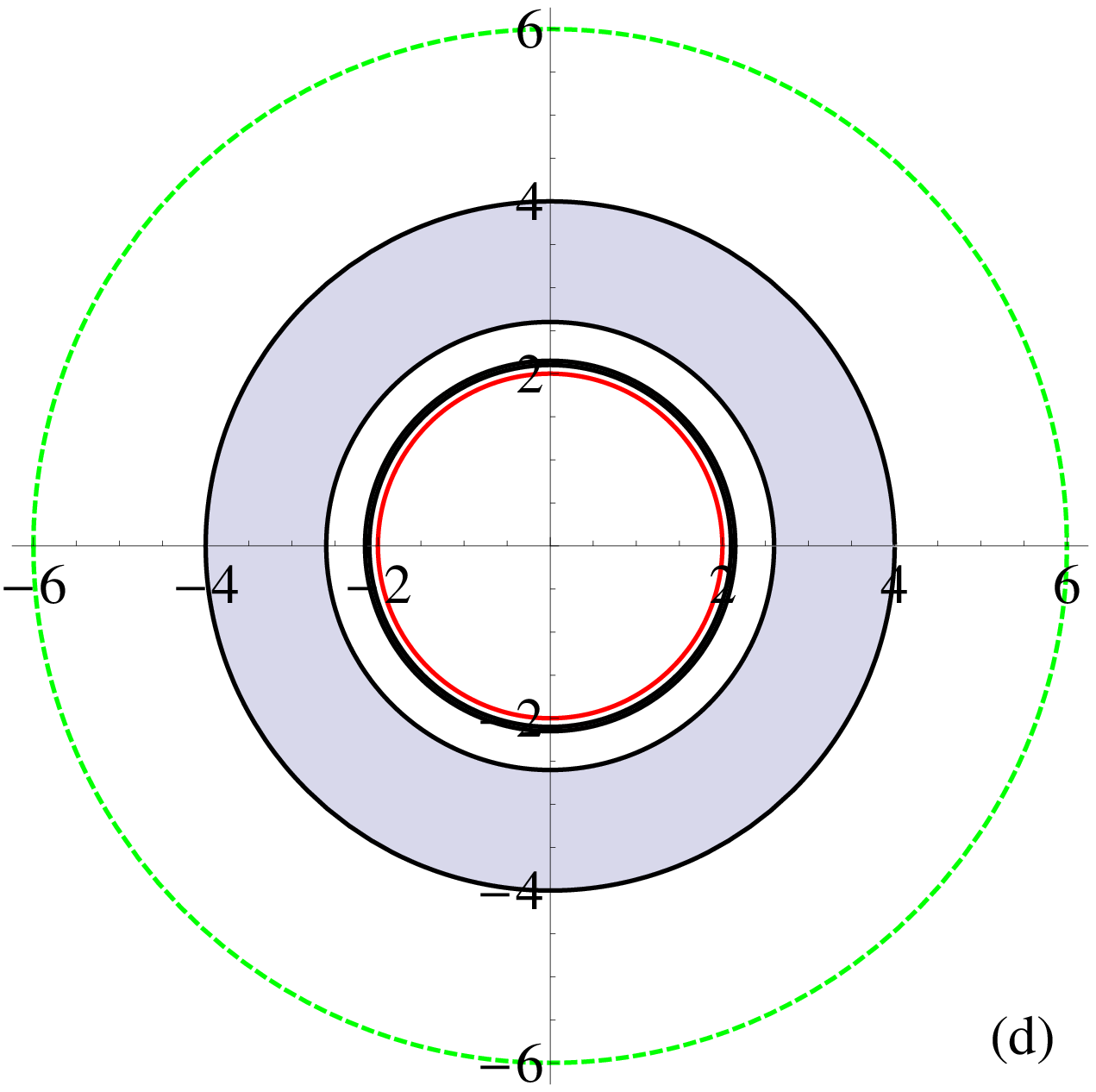}
\caption{ (a) Radius of the last stable circular orbit as a function of the quadrupole. 
Below the critical radius $r_c=m(3+2q)$, no motion is allowed. The outer singularity $r_{sing}=2m$ is also plotted. Accretion disks are illustrated for positive $q$ in 
(b),
for $-0.5>q>0$ in (c) and for $q=-0.52$ in (d). }
\label{fig1}%
\end{figure}

As a first approximation, an accretion disk can be considered as a set of test particles moving along circular orbits around 
the central mass. In this case, the geodesic equations on the equatorial plane are equivalent to the equations for the motion in the effective potential \cite{geoqmetric15}
\begin{equation}
\label{veff}
V_{eff}^2(r,q)=  \left(1-\frac{2m}{r}\right)^{q+1} \left[\frac{  l ^2}{r^2}\left(1-\frac{2m}{r}\right)^{q} +\epsilon \right] \ ,
\end{equation}
where $E$ and $l$ are constants of motion. 
The radius and stability properties of circular orbits are completely determined by the behavior of the effective potential which, in turn, depends on the behavior 
of the first and second radial derivatives. We performed a detailed analysis of the behavior of the effective potential. 
The result of this study is summarized in Fig. \ref{fig1}. The region of stability determines the spatial region where
an accretion disk can exist, and the radius of the last stable circular orbit is interpreted as the minimum inner radius of the disk.
%\begin{center}
%\end{center}

 We see that for positive values of $q$ the accretion disk is always 
located outside the Schwarzschild radius of the last stable circular orbit $r^{Sch}_{lsco}=6m$.  
For negative values of $q$, the disk can be completely inside the radius  $r^{Sch}_{lsco}=6m$. Finally, for values of $q$ close to $-0.5$, 
a second inner disk appears in a region very closed to the outer singularity $r_{sing}=2m$.   

We conclude that the quadrupole parameter can modify drastically the geometric structure of accretion disks. Therefore, it should be possible to determine the value of $q$ from 
the geometric properties of the disk.

We acknowledge the support through a Grant of the Target Program of the MES of the RK, Grant No. 3101/GF4 IPC-11/2015, 
  DGAPA-UNAM, Grant No. 113514, and Conacyt, Grant No. 166391.

%\section{Accretion disks}

%\section*{Acknowledgments}

\end{document}